\documentclass[aps,prd,twocolumn,fleqn]{revtex4-1}

\usepackage{bm}
\usepackage{graphicx}
\usepackage{amsmath}
\usepackage{calc}
\usepackage{epstopdf}

\newcommand{\be}{\begin{equation}}
\newcommand{\ee}{\end{equation}}
\newcommand{\ba}{\begin{eqnarray}}
\newcommand{\ea}{\end{eqnarray}}
\newcommand{\bd}{\begin{displaymath}}
\newcommand{\ed}{\end{displaymath}}

\def\l{\left}
\def\r{\right}

\def\thalf{{\textstyle{\frac{1}{2}}}}

\newcommand{\avg}[1]{\bigl\langle #1 \bigr\rangle_S} 
\newcommand{\mean}[1]{\bigl\langle #1\bigr\rangle} 

\begin{document}

\title{Hydrodynamic fluctuations near a critical endpoint and Hanbury Brown-Twiss interferometry}

\author{Christopher Plumberg and Joseph I. Kapusta}
\affiliation{School of Physics and Astronomy, University of Minnesota, Minneapolis, Minnesota, 55455, USA}

\begin{abstract}
The field of high energy nuclear collisions has witnessed a surge of interest in the role played by hydrodynamic fluctuations.  Hydrodynamic fluctuations may have significant effects on matter created in heavy-ion accelerators whose trajectories in the plane of temperature versus chemical potential pass near a possible critical endpoint.  We extend previous studies to explore the impact of these fluctuations on Hanbury Brown-Twiss interferometry of identical hadrons.  With an appropriately defined correlation function we find that the fluctuations increase substantially when the trajectory passes near a critical endpoint, and also displays a damped oscillatory behavior in the rapidity distance $\Delta y$ unlike that originating from initial-state fluctuations.
\end{abstract}


\date{\today}

\maketitle

\section{Introduction}
\label{Sec0}

Heavy ion collisions can be probed experimentally by means of many different kinds of physical observables, such as the anisotropic flow coefficients $v_n$ \cite{Poskanzer:1998yz}, mean transverse momentum $\mean{p_T}$ \cite{Wang:1998ww}, and the Hanbury Brown-Twiss (HBT) radii \cite{Frodermann:2006sp}.  The values assumed by these observables in any single heavy ion collision will tend to differ from their mean values taken over a collection or ensemble of such events; the observables in question are said to fluctuate randomly from event to event.  These event-by-event fluctuations can be characterized by their statistical properties in the form of their own probability distributions over the set of events, thereby providing the ability to compare theoretical predictions with experimental measurements in a systematic fashion and to place corresponding constraints upon various aspects of heavy-ion collisions and their evolution \cite{Qiu:2011hf,Niemi:2012aj,Jia:2013tja}.
	
Event-by-event fluctuations may originate in several different ways.  One of the most significant sources of fluctuations, which has received a great deal of both theoretical and experimental attention in recent years, is due to randomness in the initial state of the colliding nuclei, such as the positions of nucleons within each of the nuclei.  Another source of fluctuations which has received somewhat less attention, but which is rapidly gaining attention, is known as hydrodynamic fluctuations.  The terms hydrodynamical noise, thermal noise, and thermal fluctuations have variously been used in the literature \cite{Stephanov:2001zj,Kapusta:2012sd,Young:2013fka}.  These fluctuations, which are consequences of the fluctuation-dissipation theorem in near-equilibrium systems, can affect the hydrodynamic evolution of the produced matter throughout its evolution.  In general, these fluctuations are expected to be sub-leading to the effects of the initial-state fluctuations mentioned above, except perhaps when the evolution of the system passes near a phase transition or critical end point \cite{Gavai:2004sd, Nonaka:2004pg}.  Such a scenario is particularly relevant to the description of results from the Beam Energy Scans I and II (BESI and BESII) at the Relativistic Heavy Ion Collider (RHIC) as well as the planned program at the Facility for Antiproton and Ion Research (FAIR) \cite{Spiller:2006gj, STAR:BESI, Shigaki:2010zz, Abgrall:2014xwa}.
	
Some of the effects of hydrodynamic fluctuations have been investigated earlier \cite{Kapusta:2011gt, Kapusta:2012zb}, and we borrow heavily from those papers.  We extend that work to explore the behavior of the HBT radii in the presence of such fluctuations, while neglecting initial-state fluctuations, with 1+1 dimensional Bjorken flow.  
	
We have organized our paper in the following way.  In Sec. \ref{Sec1}, we briefly review the models for the Quantum Chromodynamics (QCD) equation of state \cite{Kapusta2010} and hydrodynamical noise \cite{Kapusta:2012zb} which allows us to study the effects of the conjectured critical endpoint in heavy ion collisions.  We then discuss in Sec. \ref{Sec2} how we embed these models into a system exhibiting Bjorken evolution, as is expected in heavy ion collisions to a limited extent.  We emphasize that much of the work presented in Secs. \ref{Sec1} and \ref{Sec2} is not original and has been discussed more completely elsewhere.  In Sec. \ref{Sec3} we extend the formalism of Secs. \ref{Sec1} and \ref{Sec2} to the calculation of the HBT radii and their event-by-event distributions in a Bjorken scenario.  In Sec. \ref{Sec4} we show how to construct physical observables which allow us to quantify the effects of hydrodynamical fluctuations on the HBT radii, and we use our model to compute these observables numerically and show that the effects of fluctuations can be appreciable in high energy nuclear collisions.  Our conclusions are presented in Sec. \ref{Sec5}.

\section{Fluctuations and the critical endpoint}
\label{Sec1}
	
Hydrodynamical noise is a generic feature of fluid dynamical systems.  It arises from treating the microscopic dynamics of the system in terms of a course-grained, macroscopic description with only a few thermodynamic variables, such as temperature and chemical potential.  The fluctuation-dissipation theorem, however, guarantees that the squared-amplitudes of these fluctuations arising from the microscopic dynamics can be related to transport coefficients present in the macroscopic description of the fluctuating systems \cite{Kubo:1966}.  Additionally, these fluctuations can have a pronounced effect on the overall evolution of the system under certain conditions, leading to a breakdown of the simplest, macroscopic description.
	
These considerations motivate a systematic incorporation of such fluctuations into the standard hydrodynamic description of evolving systems, such as are created in heavy-ion collisions.  Historically, the most common way of doing this has been to add stochastically fluctuating source terms to the usual hydrodynamic equations which describe the macroscopic evolution.  
In this case, the equations of motion become stochastic differential equations whose solutions are determined by the statistical properties of the fluctuations, as specified by the moments of the probability distributions which govern the fluctuating source terms.  For fluctuations in the baryon number it was shown in Refs. \cite{Kapusta:2011gt,Kapusta:2012zb} that the relevant correlation function in the context of the Bjorken model is
\be
\langle f(x_1) f(x_2) \rangle = 2 \lambda \left( \frac{nT}{sw}\right)^2 \delta(x_1 - x_2) \ ,
\label{bfluctuations}
\ee
Near a critical point the thermal conductivity $\lambda$ diverges.  In Ref. \cite{Kapusta:2012zb} a model for $\lambda$ was developed for the purported critical point in QCD.  We use that model in what follows.

In Refs. \cite{Kapusta2010,Kapusta:2012zb} it was assumed that the critical point lies somewhere along a crossover curve parameterized by
\be
\left(\frac{T}{T_0}\right)^2 + \left(\frac{\mu}{\mu_0}\right)^2 = 1 \ ,
\label{Xover}
\ee
with an estimate of $T_0 \approx 170$ MeV and $\mu_0 \approx 1200$ MeV.  The high energy density equation of state was taken to be of the form
\be
P = A_4 T^4 + A_2 T^2 \mu^2 + A_0 \mu^4 - C T^2 - B \ .
\label{TsquareEOS}
\ee
The coefficients $A_i$ are the same as for a noninteracting gas of massless gluons and $N_f$ flavors of massless quarks.
\ba
A_4 &=& \frac{\pi^2}{90} \left( 16 + \frac{21N_f}{2} \right) \nonumber \ , \\
A_2 &=& \frac{N_f}{18} \nonumber \ , \\
A_0 &=& \frac{N_f}{324 \pi^2} \ .
\ea
Those papers used $N_f=2$, $T_0=170$ MeV, and $\mu_0 = 1218.5$ MeV.  The numerical choice for $\mu_0$ means that the pressure is constant along the curve given by Eq. (\ref{Xover}).   Finally $B = 0.8 T_0^2$.  This equation of state is not valid at too low energy densities.  In principle it should incorporate the critical point behavior to be fully consistent with the thermal conductivity.  We will use it nevertheless in the same spirit as in \cite{Kapusta:2012zb}, namely, as a numerically tractable exploratory study of the effects of hydrodynamical fluctations on HBT correlation functions near a critical point.

\section{Bjorken expansion with hydrodynamic fluctuations}
\label{Sec2}

The energy-momentum tensor in ideal fluid dynamics is
\be
T^{\mu\nu}=wu^{\mu}u^{\nu}-Pg^{\mu\nu} \ .
\ee
The shear and bulk viscosities are set to zero to focus on the effects of thermal conductivity.  In boost-invariant hydrodynamics the flow velocity has the nonvanishing components
\ba
u^0 &=& \cosh(\xi + \omega) \nonumber \ , \\
u^3 &=& \sinh(\xi + \omega) \ . 
\ea
Here $\xi$ is the space-time rapidity and $\omega(\xi,\tau)$ is a fluctuation that depends on both $\xi$ and the proper time $\tau$.  The baryon current is 
\be
J^{\mu} = n u^{\mu} + I^{\mu} \ ,
\ee
where $I^{\mu}$ is a fluctuation.  The smooth, background fluid equations lead to the well-known simple solutions
\be 
\label{eq:sevol}
s(\tau) = s_i \tau_i/\tau
\ee
and 
\be 
\label{eq:nevol}
n(\tau) = n_i \tau_i/\tau \ ,
\ee
where $s_i$ and $n_i$ are the entropy and baryon densities at some initial time $\tau_i$. 

\begin{figure}
\centering
\includegraphics[width=0.95\linewidth]{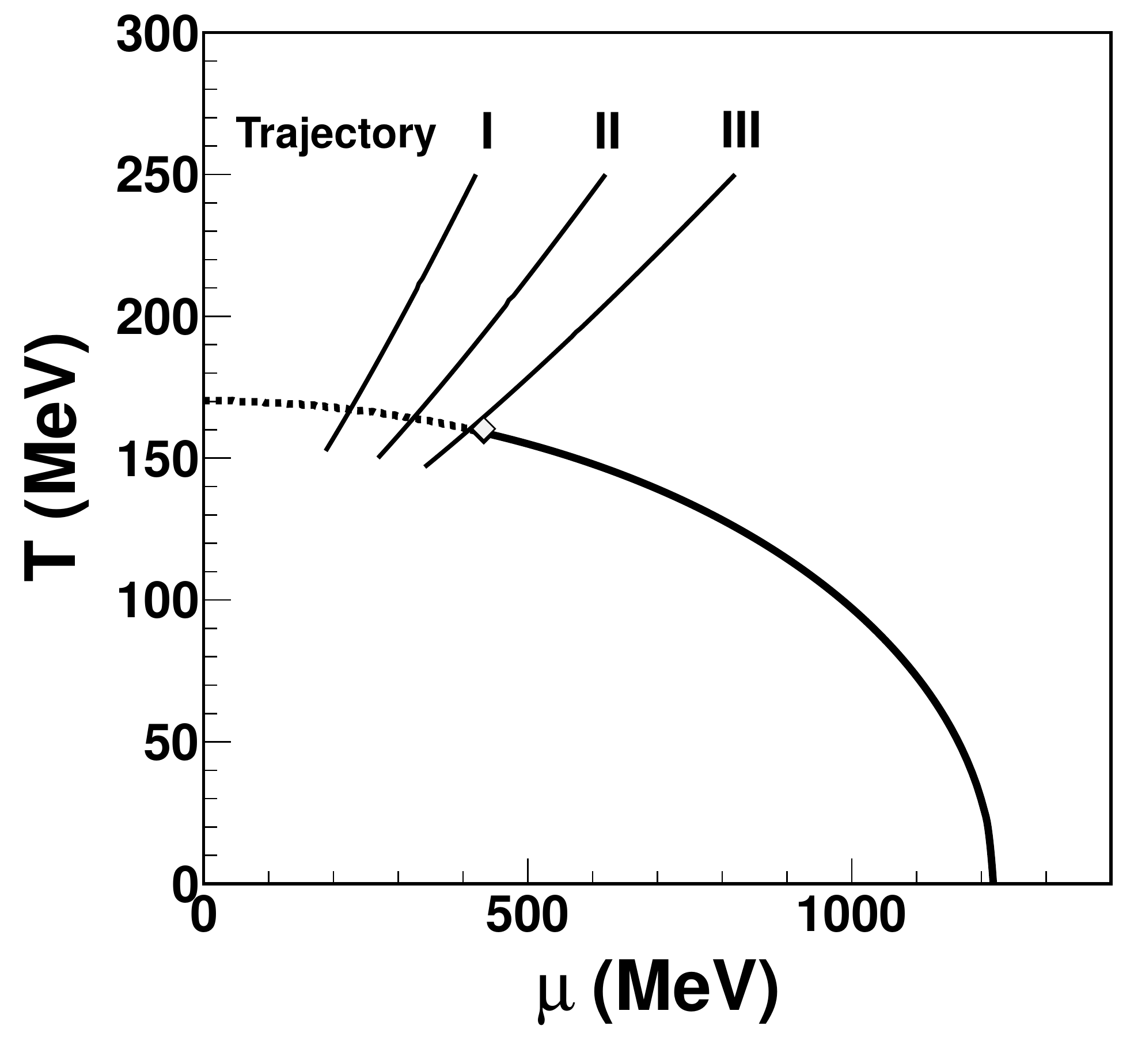}
  \caption{The phase diagram showing the crossover curve and the three trajectories used in the computation.  Taken from Ref. \cite{Kapusta:2012zb}.}
  \label{fig:trajectories}
\end{figure}

Some representative solutions for $s$ and $n$ use an initial temperature of $T_i = 250$ MeV, initial time of $\tau_i = 0.5$ fm/c, and initial chemical potentials of $\mu_i =$ 420, 620, and 820 MeV.  The adiabatic trajectories corresponding to these three cases are shown in Fig. \ref{fig:trajectories}.  Trajectories I and II represent crossover transitions while trajectory III passes very close to the critical end point, which is here chosen to be located at $T_c = 160$ MeV and $\mu_c = 411.74$ MeV.  The entropy per baryon at the critical point is 19.96, while for trajectories I, II and III it is 37.98, 26.08 and 20.06, respectively.  The time evolution is terminated when the zero pressure curve is reached, which is at $\tau_f = $ 3.0, 3.3, and 3.7 fm/c, respectively.  In reality, matching to a full hadronic equation of state should be done, but we don't do it for this illustrative example.

The full equations are linearized in the fluctuations, such as $\delta n$ and $\delta s$, and these in turn are linear functionals of $I^{\mu}$.  The nonvanishing components are
\ba
I^0 &=& s(\tau) f(\xi,\tau) \sinh \xi \nonumber \ , \\
I^3 &=& s(\tau) f(\xi,\tau) \cosh \xi
\ea  
on account of the condition that $u_{\mu} I^{\mu}=0$. Notice that $f$ is dimensionless as the entropy density has been factorized out for convenience. The average value 
$\langle f(\xi,\tau) \rangle = 0$ while the correlator was given in Eq. (\ref{bfluctuations}).   

The linearized equations are conveniently expressed in terms of the dimensionless variables $\delta s/s$, $\delta n/s$, and $\omega$.
\be
\label{eom1c} \tau \frac{\partial}{\partial \tau} \left( \frac{\delta s}{s} \right) + \frac{\partial \omega}{\partial \xi} - \frac{\mu}{T} \frac{\partial f}{\partial \xi} = 0
\ee
\be
\label{eom2c} \tau \frac{\partial}{\partial \tau} \left( \frac{\delta n}{s} \right)  + \frac{n}{s} \frac{\partial \omega}{\partial \xi} + \frac{\partial f}{\partial \xi} = 0
\ee
\be
\label{eom3c} \tau \frac{\partial \omega}{\partial \tau} + (1-v_{\sigma}^2)\omega + \frac{v_n^2Ts}{w}\frac{\partial}{\partial \xi} \left( \frac{\delta s}{s} \right)
+ \frac{v_s^2\mu s}{w}\frac{\partial}{\partial \xi} \left( \frac{\delta n}{s} \right) = 0
\ee
Here $v_{\sigma}^2$ is the physical, adiabatic speed of sound squared where the derivative is taken at fixed entropy per baryon $\sigma = s/n$, while $v_n^2 \equiv (\partial P/\partial \epsilon)_n$ and $v_s^2 \equiv (\partial P/\partial \epsilon)_s$.  They are related by 
\be
v_{\sigma}^2 = \frac{Tsv_n^2 + \mu nv_s^2}{w} \ .
\ee

These equations can be solved by Fourier transformation on the variable $\xi$ due to boost invariance.  For any function $X(\xi,\tau)$ its Fourier fransform is   
\be
\tilde{X}(k,\tau) = \int_{-\infty}^{\infty} d\xi {\rm e}^{-ik\xi} X(\xi,\tau) \,.
\ee
In the presence of the fluctuating forces, the solution to the equations of motion for $X=\delta s/s, \delta n/s,\omega$ as well as for $\delta P/(Ts)$ and $\delta \sigma$ reads
\be 
\tilde{X} (k,\tau)= - \int_{\tau_0}^{\tau} \frac{d\tau'}{\tau'} \tilde{G}_X (k,\tau,\tau') \tilde{f} (k,\tau') \ .
\ee
The response functions $\tilde{G}_X$ were explicitly given in \cite{Kapusta:2012zb}.  The correlation function of the fluctuating force in $\xi,\tau$ coordinates is
\bd 
\langle f(\tau_1, \xi_1) f(\tau_2, \xi_2) \rangle =
\ed
\be
\frac{2 \lambda(\tau_1)}{A_{\perp} \tau_1} \left[ \frac{n (\tau_1) T(\tau_1)}{s(\tau_1) w(\tau_1)} \right]^2 \delta(\tau_1 -\tau_2) \delta(\xi_1 - \xi_2) \ , 
\ee
with
\be 
\delta (x_1-x_2) = \frac{1}{A_{\perp}\tau_1} \delta(\tau_1-\tau_2) \delta(\xi_1-\xi_2) 
\ee
and where $A_{\perp}$ is the transverse area in the Bjorken model.  The Fourier transform is
\bd 
\langle \tilde{f}(k_1,\tau_1) \tilde{f}(k_2,\tau_2)\rangle = \frac{4 \pi \lambda(\tau_1)}{A_{\perp} \tau_1} \nonumber \\
\ed
\be
\times \left[ \frac{n (\tau_1) T(\tau_1)}{s(\tau_1) w(\tau_1)} \right]^2 \delta(\tau_1-\tau_2) \delta(k_1+k_2) \ . 
\ee

In Fourier space the correlation function of a pair of fluctuating variables $X$ and $Y$ is
\bd
\langle \tilde{X} (k_1,\tau_1) \tilde{Y} (k_2,\tau_2) \rangle = \frac{4 \pi}{A_{\perp}} \delta(k_1+k_2) \int_{\tau_0}^{{\rm min}(\tau_1,\tau_2)} 
\frac{d\tau}{\tau^3} \lambda(\tau) \nonumber \\
\ed
\be
\times \left[ \frac{n (\tau) T(\tau)}{s(\tau) w(\tau)} \right]^2 \tilde{G}_X (k_1;\tau_1,\tau) \tilde{G}_Y (k_2;\tau_2,\tau) \ . 
\ee
The equal-time correlator at the final or freeze out time $\tau_f$ is
\bd 
\tilde{C}_{XY} (k; \tau_f) =\frac{4 \pi}{A_{\perp}} \int_{\tau_0}^{\tau_f}  \frac{d\tau}{\tau^3} \lambda(\tau) \nonumber \\
\ed
\be
\times \left[ \frac{n (\tau) T(\tau)}{s(\tau) w(\tau)} \right]^2 \tilde{G}_{XY} (k;\tau_f,\tau) \ , 
\ee
where
\be 
\tilde{G}_{XY} (k; \tau_f,\tau) = \tilde{G}_X (k; \tau_f,\tau) \tilde{G}_Y (-k;\tau_f,\tau) \ . 
\ee
Finally, the correlation function in space-time rapidity is
\bd
C_{XY} (\xi_1-\xi_2, \tau_f) = \frac{2}{A_{\perp}} \int_{\tau_0}^{\tau_f} \frac{d\tau}{\tau^3} \lambda(\tau)  \nonumber \\
\ed
\be
\times \left[ \frac{n (\tau) T(\tau)}{s(\tau) w(\tau)} \right]^2  G_{XY} (\xi_1-\xi_2; \tau_f ,\tau) \ , 
\ee
where $G_{XY} (\xi_1-\xi_2; \tau_f ,\tau)$ is the inverse Fourier transform of $\tilde{G}_{XY} (k; \tau_f,\tau)$.  

\section{HBT fluctuations}
\label{Sec3}

Bjorken expansion is a 1+1-dimensional approximate characterization of the space-time structure of heavy-ion collisions and so necessarily fails to capture many essential characteristics.  One particular drawback of the Bjorken approach from the standpoint of interferometric analyses is the lack of structure in the transverse plane owing to the reduced dimensionality of the system.  Nevertheless, it can suffice to provide qualitative insights into how the presence of hydrodynamic fluctuations influences the interferometric radii.  Moreover, since the reduced dimensionality of the Bjorken system eliminates the outward and sideward directions, we will restrict our attention to event-by-event analyses of the longitudinal radius $R^2_l$ in such systems.
	
The strategy we adopt in this section can be outlined with a few general steps.
\begin{enumerate}
\item Define and compute particle production and the event-by-event radius $R^2_l(y)$ for the system at freeze-out at momentum rapidity $y$.
\item Expand $R^2_l(y)$ to first order in the fluctuations $\delta T$, $\delta \mu$, and $\omega$.
\item Obtain the ensemble-averaged $\mean{R^2_l(y)}$ and use this to define the event-by-event fluctuation of $R^2_l(y) \equiv \mean{R^2_l(y)} + \delta R^2_l(y)$.
\item Finally, compute the correlator $\mean{\delta R^2_l(y_1) \delta R^2_l(y_2)}$.
\end{enumerate}
	
The most direct way of describing particle production for our purposes is by means of the emission function, or Wigner density, $S(x,K)$.  The emission function is related to the single-particle spectra by
\bd
\frac{dN}{K_T dK_T d\phi dy} = \int d^4 x\, S(x,K) =
\ed
\vspace{-6mm}
\be
\kappa \, {\rm e}^{\mu/T} \int d\xi \cosh(y - \xi) m_{\perp} \exp\left[ -\frac{m_{\perp}}{T} \cosh(y-\xi-\omega)\right] \,,
\label{singleparticlespectra}
\ee
where $\kappa = d_s A_{\perp} \tau_f/(2\pi)^3$, $d_s$ is the spin degeneracy, $A_{\perp}$ the transverse area of the system, and $\tau_f$ the proper time defining the freeze-out surface.  We emphasize that here and throughout we are referring to a single species of hadron. 

Now we generalize the treatment of two-point functions for particle production in Refs. \cite{Kapusta:2011gt,Kapusta:2012zb} to include two-point functions of other observables such as the HBT radii.  Using the well-known ``source variances approximation" to compute the HBT radii, we can also write $R^2_l$ in terms of the emission function.
\bd
R^2_l(K) \equiv \avg{(z-v_L t)^2} - \avg{z-v_L t}^2 =
\ed
\vspace{-8mm}
\be
\frac{\left[ \int d^4x \, (z-v_L t)^2 S(x,K) \right]}{\int d^4x \, S(x,K)} - \frac{\left[ \int d^4x \, (z-v_L t) S(x,K) \right]^2}{\left[ \int d^4x \, S(x,K) \right]^2} 
\label{R2l_definition}
\ee
Here $z$ corresponds to the longitudinal axis of the system, $v_L = \tanh y$ is the longitudinal velocity of the particle, and we represent the averages over the emission function $S(x,K)$ by
\begin{equation}
\avg{F(x)} \equiv \frac{\int d^4 x\, F(x) S(x,K)}{\int d^4 x\, S(x,K)}.
\label{SV_definition}
\end{equation}
After changing variables from $t,z$ to $\tau,\xi$ we find that
\bd
R^2_l(K)  \left[ \int d^4x \, S(x,K) \right]^2 =
\ed
\vspace{-6mm}
\be
\tau_f^2 \kappa^2 \int^{\infty}_{-\infty} d\xi_1 \int^{\infty}_{-\infty} d\xi_2 \, g(m_{\perp}, y, \xi_1, \xi_2) s(y,\xi_1, \xi_2)
\label{numeratorR}
\ee
where
\ba
s &=&  \tanh y \left( \sinh \xi_2 \cosh \xi_1 + \sinh \xi_1 \cosh \xi_2 \right)   \nonumber \\
&-&  \tanh^2 y \cosh \xi_1 \cosh \xi_2  - \sinh \xi_1 \sinh \xi_2 \nonumber \\
&+& \thalf ( \sinh \xi_1 - \tanh y \cosh \xi_1 )^2 \nonumber \\
&+& \thalf ( \sinh \xi_2 - \tanh y \cosh \xi_2 )^2
\label{s_defn}
\ea
and
\bd
g = m_{\perp}^2 \cosh(y-\xi_1) \cosh(y-\xi_2) \, {\rm e}^{\mu_1/T_1} {\rm e}^{\mu_2/T_2}
\ed
\vspace{-7mm}
\be
\times \exp\left[ -m_{\perp} \left( \frac{\cosh(y-\xi_1-\omega_1)}{T_1} + \frac{\cosh(y-\xi_2-\omega_2)}{T_2} \right) \right] \,.
\label{g_defn}
\ee
Finally, due to the reduced dimensionality of our system, we average over the transverse momentum dependence of $R^2_l(K)$ in order to obtain results which depend only on the momentum space rapidity $y$, namely, $R^2_l(y)$.  This amounts to the integration $\int_m^{\infty} dm_{\perp}m_{\perp}$ of the numerators and denominators separately in Eq. \ref{R2l_definition}. One can show that this definition agrees with how the $K_{\perp}$ averaged $R^2_l$ is measured experimentally \cite{Plumberg:2015mxa}.  By defining $R^2_l$ according to Eq. \eqref{R2l_definition}, we have formulated this quantity in terms of the curvature of the HBT correlation function in the limit that the relative momentum between identical pairs of particles vanishes.  Thus we can think of $C_{\rm HBT}$ as a correlator of second derivatives of the event-by-event HBT correlation function at different rapidities.  

To explore the effects of hydrodynamical fluctuations on $R^2_l$, we expand Eq. (\ref{numeratorR}) to first order in the fluctuating quantities $\delta T$, $\delta \mu$, and $\omega$.  For example, the expansion of $g$ is given by
\bd
\frac{\delta g}{g} = \frac{\delta \mu_1}{T_1} + \frac{\delta \mu_2}{T_2} - \frac{\mu_1 \delta T_1}{T_1^2} - \frac{\mu_2 \delta T_2}{T_2^2}
\ed
\bd
+ \frac{m_{\perp} \omega_1}{T_1} \sinh(y-\xi_1) + \frac{m_{\perp} \omega_2}{T_2} \sinh(y-\xi_2) 
\ed
\be
+ \frac{m_{\perp} \delta T_1}{T_1^2} \cosh(y-\xi_1) + \frac{m_{\perp} \delta T_2}{T_2^2} \cosh(y-\xi_2) \,.
\ee
Therefore we can write $R_l^2(y) = \langle R_l^2(y) \rangle + \delta R_l^2(y)$ where $\delta R_l^2(y)$ has terms linear in the fluctuations $\delta T$, $\delta \mu$ and $\omega$.  Thus $\delta R_l^2(y_2) \delta R_l^2(y_1)$ is second order in those fluctuations.  The averaging can be done using the techniques outlined in Sec. \ref{Sec2} to obtain $\langle \delta R_l^2(y_2) \delta R_l^2(y_1) \rangle$.  For this purpose it is useful to note that the fluctuations $\delta T$ and $\delta \mu$ are related to $\delta s$ and $\delta n$ by
\ba 
\delta T &=& \frac{\chi_{\mu \mu}\delta s - \chi_{T \mu} \delta n}{\Delta}  \ , \nonumber \\
\delta \mu & = & \frac{\chi_{T T} \delta n  - \chi_{T \mu} \delta s}{\Delta}\ , 
\ea
where $\chi_{TT} = \partial^2P(T,\mu)/\partial T^2$,  $\chi_{T\mu} = \partial^2P(T,\mu)/\partial T\partial \mu$, $\chi_{\mu\mu} = \partial^2P(T,\mu)/\partial \mu^2$, and 
$\Delta = \chi_{TT} \chi_{\mu \mu} - \chi_{T \mu}^2$.  More details may be found in the appendix.

\section{Numerical results}
\label{Sec4}

In this section we present the effects of hydrodynamical noise on the HBT radii near a critical point, as computed from the preceding formalism.  To begin it is useful to look at the mean square HBT radius from Eqs. (\ref{numeratorR}-\ref{g_defn}).  After shifting the variables of integration $\xi_i \rightarrow \xi_i + y$ we find that
\be
s \rightarrow \frac{(\sinh \xi_1 - \sinh \xi_2 )^2}{2 \cosh^2y} \,.
\ee
Then it is easy to show that
\be
\langle R_l^2(y) \rangle = \frac{\langle R_l^2(0) \rangle}{\cosh^2y} \,.
\ee
This dependence on $y$ is to be expected.  It just means that to an observer moving with rapidity $y$ relative to the proper frame of the HBT radius, that radius is Lorentz contracted by the Lorentz factor of $\gamma = \cosh y$.  To obtain a dimensionless correlation function, and to divide out the trivial Lorentz contraction factors, we divide by the product of the mean HBT radii.  We also multiply by $\langle dN/dy \rangle$ in order to cancel the dependence on the transverse area $A_{\perp}$ which always enters in the Bjorken model.  (Remember that   $\langle dN/dy \rangle$ is the number of particles per unit rapidity of the species under consideration; it is not the charged particle rapidity distribution, for example.)  Hence we define the correlation function
\be
C_{\rm HBT}(\Delta y)  = \l< \frac{dN}{dy} \r> \frac{\langle \delta R_l^2(y_2) \delta R_l^2(y_1) \rangle}
{\langle R_l^2(y_2) \rangle \langle R_l^2(y_1) \rangle} \,.
\label{CHBT_vs_Deltay}
\ee
The magnitude of the two-point function for $R^2_l(y)$ therefore depends on the event-averaged multiplicity, with the correlations stronger in small systems (such as $p+p$ collisions) than in large systems (such as central $Au+Au$ collisions) \cite{Yan:2015lfa}.

\begin{figure}
\centering
\includegraphics[width=0.95\linewidth]{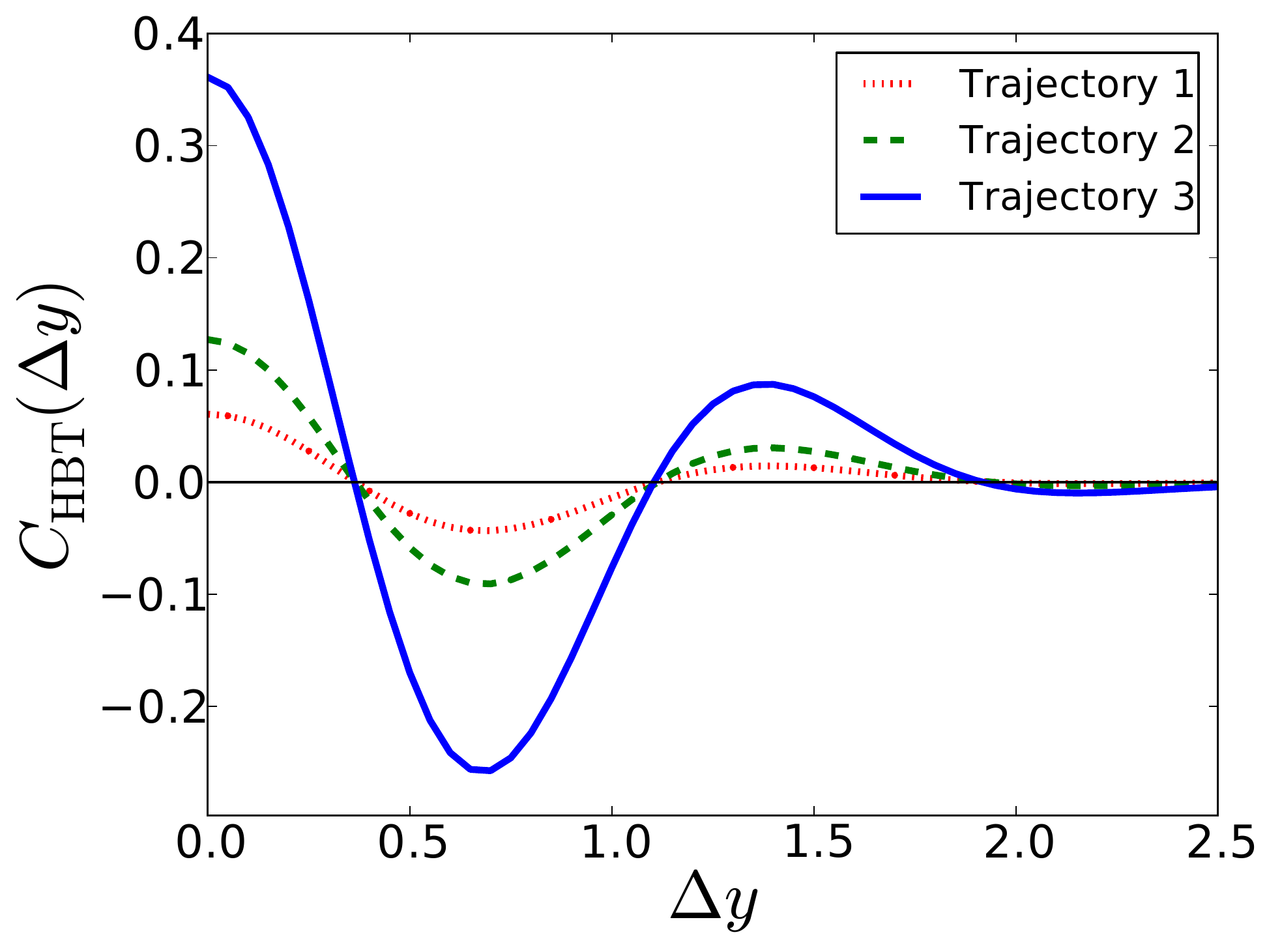}\\
\caption{Two-point HBT correlation $C_{\rm HBT}$ as a function of the rapidity difference $\Delta y$ for each of the three trajectories shown in Fig. 1 using proton pairs to define $R_l^2$.}
\label{Fig2}
\end{figure}

\begin{figure}
\centering
\includegraphics[width=0.95\linewidth]{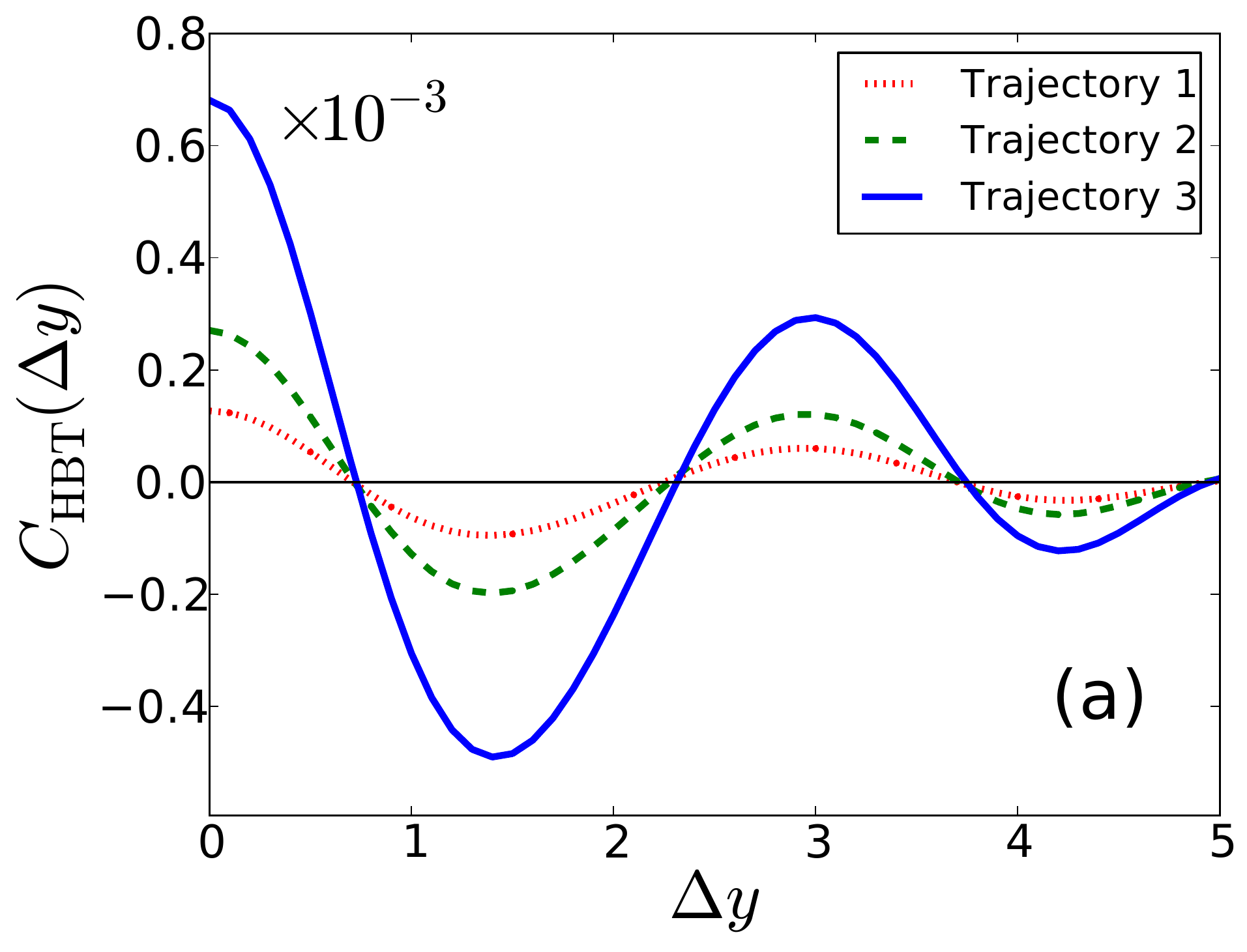}\\
\includegraphics[width=0.95\linewidth]{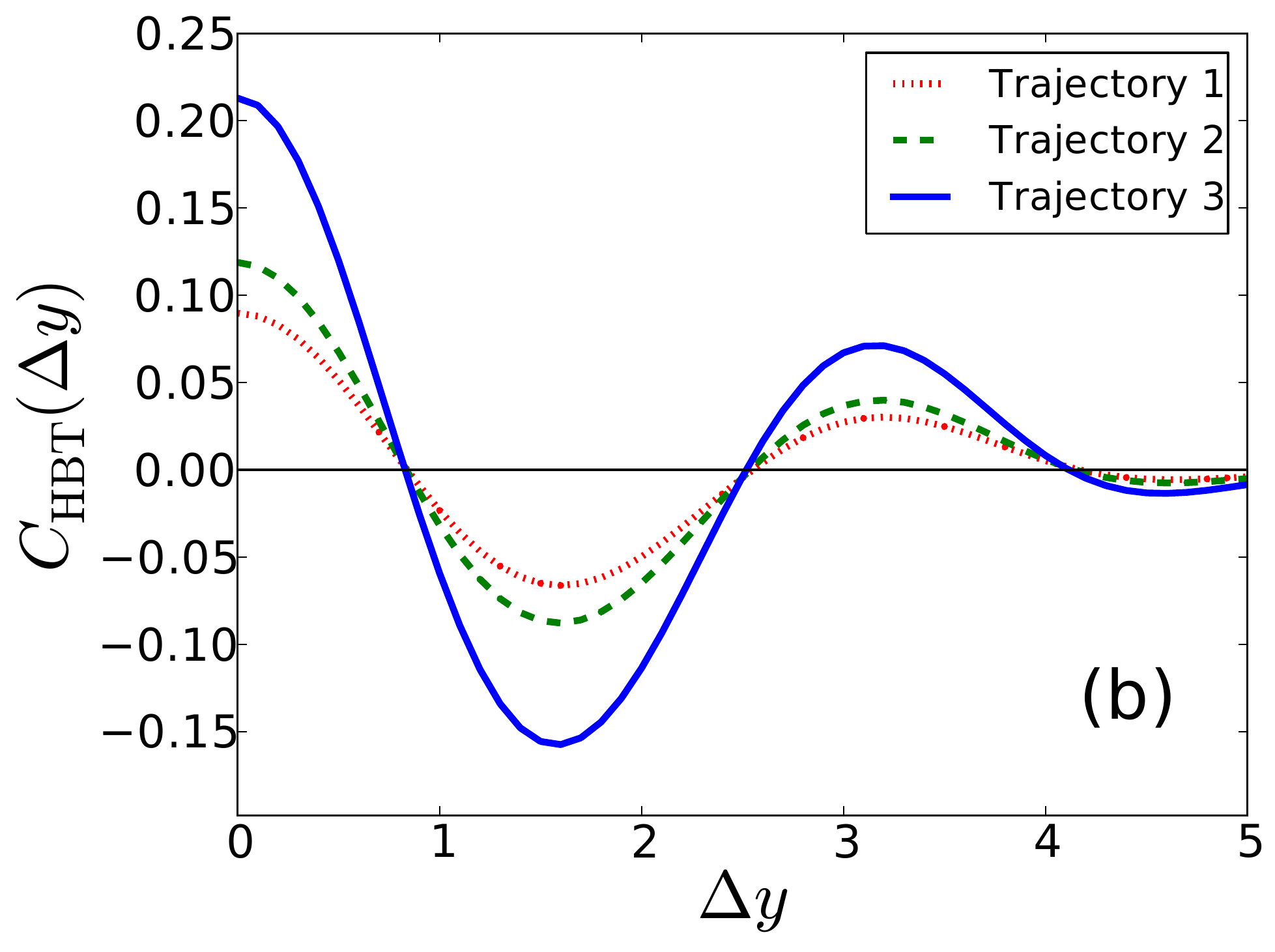}\\
\caption{Two-point HBT correlation $C_{\rm HBT}$ as a function of the rapidity difference $\Delta y$ for each of the three trajectories shown in Fig. 1 using $\pi^+$ pairs to define $R_l^2$.  The average pion chemical potential is zero.  In panel (a) we assume that it does not fluctuate, while in panel (b) we assume that it fluctuates the same as the proton, or baryon, chemical potential.}  
\label{Fig3}
\end{figure}

In Figs. \ref{Fig2} and \ref{Fig3}, we plot the correlator \eqref{CHBT_vs_Deltay} for each of the three trajectories shown in Fig. 1: in Fig. \ref{Fig2} for proton pairs, and in Fig. \ref{Fig3} for charged pion (either $\pi^+$ or $\pi^-$) pairs.  Since we have not taken into account the electric charge current the average chemical potential for pions is zero.  To estimate the effect of the inclusion of fluctuations in the pion chemical potential, as was done in Ref. \cite{Kapusta:2012zb}, we show in Fig. \ref{Fig3} the correlator both without (panel (a)) and with (panel (b)) fluctuations of the chemical potential $\delta \mu$ assuming that the latter is the same as for fluctuations in the baryon chemical potential.  The difference is quite consequential.

The strength of the correlations clearly increases with the proximity of the trajectory to the critical endpoint.  Similar to the two-particle correlation functions shown in Figs. 7 and 8 of Ref. \cite{Kapusta:2012zb}, we observe that the correlators in Figs. \ref{Fig2} and \ref{Fig3} exhibit oscillatory behavior in the rapidity difference.  As in the two-particle correlation functions, the values of $\Delta y$ for which the correlation vanishes appear to be universal, namely, approximately independent of trajectory, although it does depend on the mass of the particle.

The dependence of the correlation strength and the universality of the correlation zeros with respect to trajectory agree with the results which were obtained in \cite{Kapusta:2012zb}.  The correlations are the strongest for trajectory III which passes the closest to our critical endpoint, while the strength of the correlations for $\pi^+$s are dominated by fluctuations of the chemical potential.  With fluctuations of $\mu$ included, the HBT correlations for both protons and pions appear to be strong enough for experimental observation.

We can make some simple estimates of the observability of these fluctuations. Within the context of this model, the average $\pi^+$ multiplicity per unit rapidity per unit transverse area is $\mean{dN_{\pi^+}/dy} / A_{\perp} \sim 0.255$ fm$^{-2}$ for trajectory III (other trajectories produce similar values).  A reasonable estimate of the transverse area for central collisions is $A_{\perp} = \pi r_0^2 A^{2/3}$, with $A$ the atomic number of the nuclei and $r_0 = 1$ fm.  Using trajectory III in Fig. 3(b) the relative width of fluctuations in $R^2_l$ at $\Delta y = 0$ for pions is
\be
\frac{\sigma_l}{\mean{R^2_l}} \equiv \sqrt{\frac{\l< \r( \delta R^2_l(0) \l)^2 \r>}{\mean{R^2_l(0)}^2}} 
= \sqrt{\frac{C_{HBT}(\Delta y = 0)}{\mean{dN/dy}}} \sim 0.5 A^{-1/3} \,.
\ee
For protons, and for trajectory III, we find that $\mean{dN_{p}/dy} / A_{\perp} \sim 0.07$ fm$^{-2}$.  From Fig. 2 we get
\be
\frac{\sigma_l}{\mean{R^2_l}} \sim 1.3 A^{-1/3} \,.
\ee
For Au + Au collisions, $\sigma_l/\mean{R^2_l} \sim 9\%$ for pion radii and $\sigma_l/\mean{R^2_l} \sim 22\%$ for proton radii.  However, we must remember that this result for pions assumes that the fluctuations in the pion chemical equals that of the baryon chemical potential, which may be an incorrect assumption.

The reason for denoting these estimates with the notation $\sigma_l/\mean{R^2_l}$ is to suggest a natural comparison with the earlier work presented in \cite{Plumberg:2015aaw, Plumberg:2016sig}, which estimated the same quantity in the context of initial state fluctuations in $A+A$ collisions.  We base our numerical estimates on the intercept values of the solid blue curves in Figs.~\ref{Fig2} and \ref{Fig3}(b) (corresponding to Trajectory III) and emphasize that they are highly dependent upon the simplified model of Bjorken scaling that we are using in this work.  Nevertheless, these estimates can provide some guidance for the expected size of noise effects in different sized systems.  This is particularly relevant for the BES II, since $R^2_l$ fluctuations in the vicinity of a QCD critical point could be comparable in scale to the fluctuations estimated on the basis of initial state fluctuations  in $A+A$ collisions \cite{Plumberg:2015aaw, Plumberg:2016sig}, while the fluctuations in $p+p$ collisions can fluctuate in the same region of the phase diagram up to the size of the collision system itself.  These estimates therefore suggest that event-by-event fluctuations of the HBT radii could be elevated to the role of a powerful tool in probing the QCD phase diagram.

\section{Conclusions}
\label{Sec5}

In this paper we explored the effects of hydrodynamical noise on the HBT radii (specifically $R^2_l$) in heavy-ion collisions described by Bjorken evolution.  Following the general approach laid out in \cite{Kapusta:2011gt, Kapusta:2012zb}, we proposed a two-point correlation function to characterize event-by-event fluctuations of the longitudinal radius $R^2_l$, and computed it for both proton and pion pairs.
	
We have used these results to obtain simple numerical estimates of the order of magnitude of the $R^2_l$ correlations in heavy-ion collisions.  Our findings suggest that the effects of hydrodynamical noise on the HBT radii could become comparable to or even greater than the effects of initial-state fluctuations in the vicinity of the critical point, making the HBT radii and their event-wise distirbutions potentially useful tools in probing critical behavior at the BES II.
	
\section*{Acknowledgements}

We thank Juan Torres-Rincon for discussions.  This work was supported by the US Department of Energy (DOE) Grant DE-FG02-87ER40328.

\appendix

\onecolumngrid

\section{Calculation of $C_{\rm HBT}(\Delta y)$}
\label{Appendix:A}
In this appendix we summarize the main results needed to compute the correlator $C_{\rm HBT}(\Delta y)$ as plotted in Figs. \ref{Fig2} and \ref{Fig3}.  The numerator of this expression may be written as
\begin{equation}
\l< \delta R^2_l(y_1) \delta R^2_l(y_2) \r> =  \frac{\alpha^2_0 \tau_f^4}{ \cosh^2 y_1 \cosh^2 y_2} \int^{\infty}_{-\infty}dk\, {\rm e}^{i k \Delta y} \sum_{X,Y} \tilde{C}_{XY}(k) \tilde{\bar{F}}_X(-k) \tilde{\bar{F}}_Y(k) \,.
\label{CHBT_vs_Deltay_numerator}
\end{equation}
Here
\begin{equation}
\tilde{\bar{F}}_X(k) = \int^{\infty}_{-\infty} d\xi_1 \int^{\infty}_{-\infty} d\xi_2 \, {\rm e}^{-i k \xi_1} \l[ \l( \sinh \xi_1 - \sinh \xi_2 \r)^2  - \gamma_0 \r] \cosh(\xi_1) \cosh(\xi_2)  h_X(\xi_1, \xi_2) \, ,
\label{CHBT_vs_Deltay_numerator_defs}
\end{equation}
\begin{equation}
\alpha_0 = \l( \int^{\infty}_{-\infty} d\xi_1 \int^{\infty}_{-\infty} d\xi_2 \, \cosh(\xi_1) \cosh(\xi_2)  h_1(\xi_1, \xi_2) \r)^{-1} \, ,
\end{equation}
\begin{equation}
\gamma_0	= \alpha_0 \int^{\infty}_{-\infty} d\xi_1 \int^{\infty}_{-\infty} d\xi_2 \, \l( \sinh \xi_1 - \sinh \xi_2 \r)^2 \cosh(\xi_1) \cosh(\xi_2)  h_1(\xi_1, \xi_2)\, ,
\end{equation}
\begin{eqnarray}
\zeta(\xi_1, \xi_2) & = & \frac{T_f}{\cosh \xi_1 + \cosh \xi_2} \, , \\
h_1(\xi_1, \xi_2) & = & \int^{\infty}_m dm_{\perp} m^3_{\perp} {\rm e}^{- m_{\perp}/\zeta(\xi_1, \xi_2)}
= \zeta^4(\xi_1, \xi_2) \Gamma\l( 4, \frac{m}{\zeta(\xi_1, \xi_2)} \r) \, , \\
h_s(\xi_1, \xi_2) & = & T_f^{-2} \int^{\infty}_m dm_{\perp} m^3_{\perp} {\rm e}^{- m_{\perp}/\zeta(\xi_1, \xi_2)} \l[ \tilde{\chi}_{\mu\mu}\l( m_{\perp} \cosh(\xi_1) - \mu_f \r) - \tilde{\chi}_{T \mu} T_f \r] \nonumber\\
& = & T_f^{-2} \l[ \tilde{\chi}_{\mu\mu} \cosh(\xi_1) \zeta^5(\xi_1, \xi_2) \Gamma\l( 5, \frac{m}{\zeta(\xi_1, \xi_2)} \r) \r. \, \nonumber\\
&& \l. - \l( \tilde{\chi}_{\mu\mu} \mu_f + \tilde{\chi}_{T \mu} T_f \r)\zeta^4(\xi_1, \xi_2) \Gamma\l( 4, \frac{m}{\zeta(\xi_1, \xi_2)} \r) \r] \, , \\
h_\omega(\xi_1, \xi_2) & = & \int^{\infty}_m dm_{\perp} m^3_{\perp} {\rm e}^{- m_{\perp}/\zeta(\xi_1,\ xi_2)} \frac{m_{\perp}}{T_f} \sinh (\xi_1) \\
& = & T_f^{-1} \sinh (\xi_1) \zeta^5(\xi_1, \xi_2) \Gamma\l( 5, \frac{m}{\zeta(\xi_1, \xi_2)} \r) \, , \\
h_n(\xi_1, \xi_2) & = & \int^{\infty}_m dm_{\perp} m^3_{\perp} {\rm e}^{- m_{\perp}/\zeta(\xi_1, \xi_2)} \l[ \tilde{\chi}_{T T} T_f - \tilde{\chi}_{T \mu}\l( m_{\perp} \cosh(\xi_1) - \mu_f \r) \r]\\
& = & T_f^{-2} \l[ \l( \tilde{\chi}_{T \mu} \mu_f + \tilde{\chi}_{T T} T_f \r)\zeta^4(\xi_1, \xi_2) \Gamma\l( 4, \frac{m}{\zeta(\xi_1, \xi_2)} \r) \r. \nonumber\\
&& \l. - \tilde{\chi}_{T \mu} \cosh(\xi_1) \zeta^5(\xi_1, \xi_2) \Gamma\l( 5, \frac{m}{\zeta(\xi_1, \xi_2)} \r) \r] \,.
\end{eqnarray}
Finally, $\Gamma(n,z)$ is the incomplete Gamma function.  Using this notation we can write the ensemble averaged $R^2_l$ as
\begin{equation}
\mean{R^2_l(y)} = \frac{\tau_f^2 \gamma_0}{2 \cosh^2 y}
\end{equation}
and the ensemble-averaged rapidity spectrum as given in Eq. (88) of Ref. \cite{Kapusta:2012zb}.
	

\end{document}